\documentclass[]{spie}  %>>> use for US letter paper
\usepackage{amsmath,amsfonts,amssymb}
\usepackage{graphicx}
\usepackage[colorlinks=true, allcolors=blue]{hyperref}

\title{Wavefront Error Recovery and Companion Identification with the James Webb Space Telescope}

\author[1,2]{Matthew De Furio}
\author[3]{Marie Ygouf}
\author[4]{Alexandra Greenbaum}
\author[3]{Graça Rocha}
\author[5]{Michael Meyer}
\author[3,6]{Charles Beichman}
\author[7]{Jorge Llop-Sayson}
\author[3]{Gael Roudier}
\author[8]{Steph Sallum}
\author[9]{Jarron Leisenring}
\author[10,11,12]{Anand Sivaramakrishnan}

\affil[1]{Department of Astronomy, University of Texas at Austin, Austin, TX 78712, USA}
\affil[2]{NSF Astronomy and Astrophysics Postdoctoral Fellow}
\affil[3]{Jet Propulsion Laboratory, California Institute of Technology, Pasadena, CA 91109, USA}
\affil[4]{IPAC, Caltech, 1200 E. California Blvd., Pasadena, CA 91125, USA}
\affil[5]{Department of Astronomy, University of Michigan, Ann Arbor, MI 48109, USA}
\affil[6]{NASA Exoplanet Science Institute, Infrared Processing and Analysis Center (IPAC)}
\affil[7]{California Institute of Technology, 1200 E. California Blvd., Pasadena, CA 91125, USA}
\affil[8]{Department of Astronomy and Astrophysics, University of California, Santa Cruz, CA 95064, USA}
\affil[9]{Steward Observatory, University of Arizona, Tucson, AZ 85721, USA}
\affil[10]{Space Telescope Science Institute, 3700 San Martin Drive, Baltimore, MD 21218, USA}
\affil[11]{Astrophysics Department, American Museum of Natural History, 79th Street at Central Park West, New York, NY 10024}
\affil[12]{Department of Physics and Astronomy, Johns Hopkins University, 3701 San Martin Drive, Baltimore, MD 21218, USA}

\authorinfo{Further author information: (Send correspondence to M.D.F.)\\M.D.F.: E-mail: defurio@utexas.edu}

\begin{document} 
\maketitle

\begin{abstract}
The James Webb Space Telescope is orders of magnitude more sensitive than any other facility across the near to mid-infrared wavelengths. Many approved programs take advantage of its highly stable point spread function (PSF) to directly detect faint companions using diverse high-contrast imaging (HCI) techniques. However, periodic re-phasing of the Optical Telescope Element (OTE) is required due to slow thermal drifts distorting to the primary mirror backplane along with stochastic tilt events on individual mirror segments. Many programs utilize observations of a reference star to remove the stellar contribution within an image which can typically take half of the total allocated time. We present a high-contrast imaging technique for the NIRISS instrument that uses the measured wavefront error (WFE) from a phase calibration observation (performed roughly every 48 hours) as prior information in a Bayesian analysis with nested sampling. This technique estimates the WFE of a given observation and simultaneously searches for faint companions, without using a reference star. We estimate the wavefront error for both full aperture and aperture masking interferometry (AMI) imaging modes using three low order Zernike coefficients per mirror segment, using the Hexike basis, to generate synthetic PSFs and compare to simulations. We compare our technique to traditional interferometric analysis in realistic NIRISS F430M simulations both relative to the photon noise limit, and through recovering an injected companion with $\Delta$F430M= 8 mag at 0.2''. With future testing, this technique may save significant amounts of observing time given the results of our current implementation on NIRISS simulations.
 
\end{abstract}


\section{INTRODUCTION}
\label{sec:intro}  % \label{} allows reference to this section

Thousands of exoplanets have been discovered over the last twenty years, revealing a large diversity of properties \cite{Gaudi2021exbi.book....2G}.  These planets span orders of magnitude in mass, radius and semi-major axis, and exhibit varied atmospheres.  Exoplanets are most commonly detected using the radial velocity method \cite{Mayor1995Natur.378..355M} and the transit method \cite{Charbonneau2000ApJ...529L..45C}.  However, these methods are indirect and typically sensitive at separations $<$ 10 au.  Direct imaging is another observing technique that can detect exoplanets, and is most sensitive at wide separations where high contrast is achievable \cite{Bowler2016PASP..128j2001B}.  Through the direct detection of exoplanets, we can test the accuracy of theoretical models, obtain spectra to characterize atmospheres, and constrain fundamental models of planet formation.  With the launch of the \textit{James Webb Space Telescope} (JWST) \cite{Gardner2006SSRv..123..485G}, many programs will utilize high contrast imaging techniques over 0.6 - 28 $\mu$m to explore wide separations, sensitive to planetary masses previously unattainable from the ground.

Many observing and post-processing techniques have been developed to increase sensitivity to planets by removing the stellar contributions present within the image.  Reference differential imaging (RDI) involves observations of an assumed single star close in time to the target of similar spectral type, followed by subtraction of the two to identify a faint companion in the residuals \cite{Lafreniere2009ApJ...694L.148L}. Ground-based applications of this technique are sensitive to changes in the atmosphere and works best in scenarios with stable telescope optics and point-spread function (PSF). Another approach to high-contrast imaging (angular differential imaging, ADI) takes advantage of angular diversity. From the ground, ADI involves many integrations of a target over an extended period of time with the telescope rotator turned off to allow for motion of the field of view due to the Earth's rotation \cite{Marois2006ApJ...641..556M}.  This causes a potential companion to rotate in the image while keeping quasi-static speckles of the star at the same position. ADI is also applied to space-based imaging where the target is observed at multiple angles after the roll of the telescope, also referred to as roll subtraction.  Principal component analysis (PCA) is a post-processing technique applied to various high-contrast imaging data sets which attempts to reconstruct the PSF of an observation fitting an orthogonal basis set to the images determined by variation in the data and estimating the component coefficients within each integration \cite{Jee2007PASP..119.1403J, Amara2012MNRAS.427..948A, Soummer2012ApJ...755L..28S}. However, these post-processing techniques are difficult to perform on small angular scales ($<$ 1'') due to either the physical size of a coronagraphic mask, the dominance of residual starlight, or the evolution of speckles in ground-based imaging on a faster scale than the time required for the rotation to translate to motion of one full width at half maximum \cite{Marois2006ApJ...641..556M}.

Aperture masking interferometry (AMI) \cite{Haniff1987Natur.328..694H, Tuthill2000PASP..112..555T} is an observing mode that utilizes a pupil mask to convert the telescope into a non-redundant interferometric array where the interference of the light from each sub-aperture produces an interferogram from which interferometric observables can be extracted. Analyses are then applied to the interferometric observables (e.g. closure phases, squared visibilities) after calibration from reference star observations. Kernel-phase interferometry (KPI) \cite{Martinache2010ApJ...724..464M} is a similar interferometric analysis but applied to full aperture direct images. With knowledge of the telescope pupil, models in the interferometric observables are explored to find a signal from the planet that deviates from the expectations of a point source (e.g. closure phases equal to zero, squared visibilities equal to one). These techniques are most sensitive to companions at close separations.

Space-based imaging offers significant advantages over ground-based imaging. The PSF is stable due to the location of the telescope above the atmosphere. JWST is ideal for directly imaging exoplanets where it has a significantly larger diameter than previous infrared telescopes (e.g. 0.85m diameter of Spitzer) and there is low background emission in the thermal infrared. JWST is unique compared to other space-based telescopes in that it has 18 hexagonal mirror segments that define the primary mirror. Within the pupil and filter wheels of NIRCam, three weak lenses of -8, +4, and +8 waves of defocus exist and two are used to inject a specified amount of defocus roughly every two days for wavefront sensing and control (WFSC) operations to measure the integrity of the PSF \cite{Perrin2016SPIE.9904E..0FP}. The optical path difference (OPD) is extracted from these observations, and then evaluated to determine any required fine-phasing due to changes in the OPD over time, correcting for large rms residuals when appropriate using actuators on each mirror segment. Small wavefront drifts occur gradually due to minor changes in the equilibrium temperature \cite{Perrin2018SPIE10698E..09P, McElwain2023PASP..135e8001M} which will be realized over the two days of observing between WFSC operations (e.g. 8-50 nm) \cite{Rigby2023PASP..135d8001R}. Even large mirror segment tilts can occur due to the release of stress in the backplane of the primary mirror.

Many high-contrast imaging programs observe a reference star directly before or after the target star in order to perform RDI and minimize the amount of residual WFE between observations \cite{Hinkley2022PASP..134i5003H}.  Reference stars are chosen based on their proximity to the target and with similar color and brightness as the target in order to ensure a similar PSF, requiring the total charged time to be roughly twice as long as time on target. Instead, it is possible to use synthetic reference PSFs as models in post-processing techniques due to the optical stability of JWST, its slowly varying PSF, and frequent OPD measurements \cite{Greenbaum2023ApJ...945..126G}. Additionally, synthetic PSF models constructed from the measured OPD can be used as prior information to inform a fitting routine that accounts for WFE drifts through modeling the instrumental OPD to estimate both the signal and the wavefront simultaneously \cite{Ygouf2013A&A...551A.138Y, Ygouf2016SPIE.9904E..5MY, Cantalloube2018arXiv181204312C, Ygouf2020SPIE11443E..3NY}.

In this paper, we investigate the WFE estimation technique with JWST imaging simulation tools where we construct synthetic PSFs from an instrument model (informed by simulated WFS OPDs) and compare to simulated data in order to estimate the real OPD used in constructing the simulation.  We simultaneously search for faint companions within the simulated data and evaluate our ability to detect them relative to other post-processing techniques.  Our method uses publicly available OPD maps and does not require the use of a reference star, significantly reducing the total charged time for a high contrast imaging program.

In Section 2, we describe the process of constructing simulated data and our estimation routine.  In Section 3, we present the achievable contrast of this technique for both full pupil and AMI imaging modes, compare them to other post-processing techniques, and demonstrate the ability to recover an injected faint companion.  In Section 4, we discuss the utility of this method and its practical application, as well as potential future development.  In Section 5, we summarize our conclusions.

\section{Methods} \label{sec:methods}

Our method requires the knowledge of JWST's optics and the wavefront error associated with the optical telescope element (OTE) measured during WFSC operations.  Because the OTE does not experience major changes in wavefront error, we can use this information as priors to perform a Bayesian analysis with nested sampling to estimate the wavefront error during the observation while simultaneously searching for a companion and estimating its position and flux.

\subsection{Simulations} \label{subsec:simulation}
First, we describe the simulation tools used to produce NIRISS images, the components of the WFE that define the specific structure of a simulated PSF, and our approach to simulating a real JWST observing program.

\subsubsection{Tools} \label{subsec:real_sims}

\begin{figure} [ht]
\begin{center}
\begin{tabular}{c} %% tabular useful for creating an array of images 
\includegraphics[height=7cm]{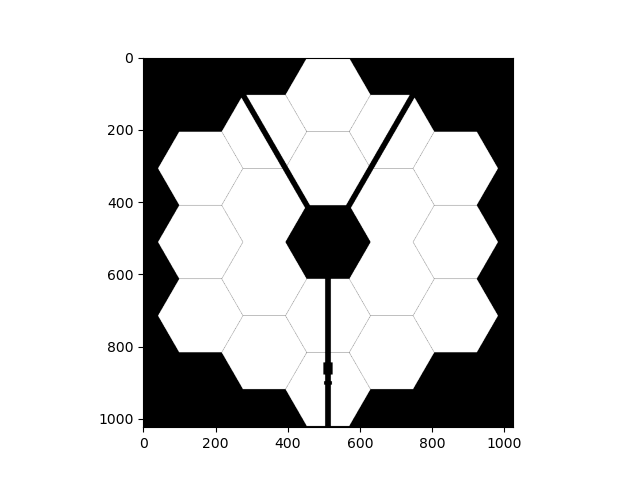}
\includegraphics[height=7cm]{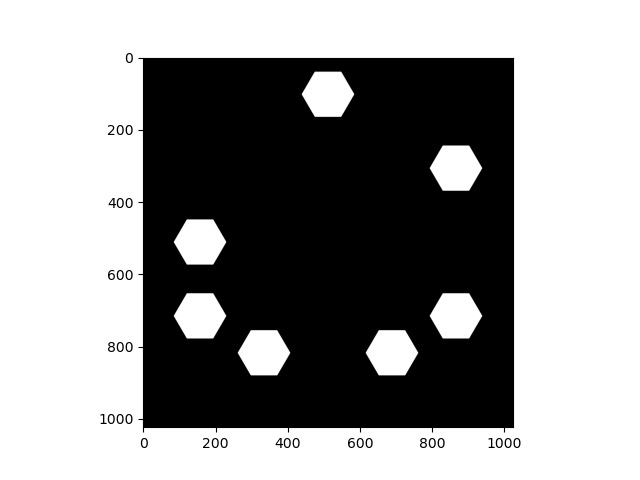}
\end{tabular}
\end{center}
\caption[pupils]
%>>>> use \label inside caption to get Fig. number with \ref{}
{ \label{fig:pupils} 
Input transmission to POPPY for the full pupil (left) and AMI (right) observing modes. X and y units are pixels.}
\end{figure} 

POPPY (Physical Optics Propagation in PYthon) \cite{Perrin2012SPIE.8442E..3DP} is a python package that can be used to simulate PSFs of a given optical system.  It requires user defined inputs that characterize the pupil geometry, the primary mirror diameter, the wavefront error for the observation, detector characteristics, and the desired field of view and sampling of the constructed image.  It is used within WebbPSF \cite{Perrin2014SPIE.9143E..3XP} to simulate realistic expectations of JWST images.

We use POPPY to produce a noiseless PSF for both the full pupil and AMI observing modes for NIRISS. We define the OPD as a combination of the prelaunch predicted OTE OPD with an added correction, specified by a Zernike polynomial expansion to describe the OPD at a field point in NIRISS, both of which are available through WebbPSF. Zernike polynomials are a set of complete orthogonal basis functions on the two-dimensional unit disk.  Each Zernike polynomial describes a specific type of optical phase aberration that can contribute to the WFE (e.g. tip/tilt, astigmatism, defocus, coma).  Each type of aberration can be more or less prominent depending on the position offset of the science instrument (SI) within the observatory focal plane and the location on the detector where a target is observed. POPPY contains a function that converts Zernike coefficients into an OPD map, which we use before combining the SI and OTE contributions of WFE.  Then, we import the transmission corresponding to each observing mode, i.e. JWST pupil or aperture mask, see Fig. \ref{fig:pupils}.  Lastly, we define the details of the detector and the image we require for our program: plate scale of NIRISS (0.065''/pixel), field of view (4'' x 4''), sampling factor (5x), and wavelength (4.2784 $\mu$m, the effective wavelength in the F430M filter). With this information, we can then produce a noiseless PSF model for NIRISS.

We import this oversampled PSF model into \texttt{ami$\_$sim} \footnote{\url{https://github.com/agreenbaum/ami_sim}}, a python package intended to simulate realistic data from NIRISS with detector level noise effects (Thatte et al., private communication).  \texttt{ami$\_$sim} injects typical appropriate noise sources including photon noise, dark current, read noise, and background. It also simulates the flat field error, as well as incorporating pointing error through sub-pixel shifts between integrations and Gaussian smoothing to represent small motion during the integration.  These pointing errors are set to 1 milli-arcsecond (mas) as is the measured value on JWST from commissioning \cite{Rigby2023PASP..135d8001R}. Expected noise effects not included are 1/f noise, inter-pixel capacitance, charge migration, and the brighter/fatter effect.

We then define the number of groups per integration, number of integrations used in the simulation, and the electron count rate in order to simulate the desired observing program.  The count rate is derived from the JWST Exposure Time Calculator (ETC) using the same observing setup as \texttt{ami$\_$sim} and scaled to the desired magnitude and spectrum of the target.  \texttt{ami$\_$sim} convolves the input oversampled PSF with a user-defined scene to produce the detector-sampled image, e.g. single star, star with faint companion, etc.  \texttt{ami$\_$sim} then produces a data cube of the astrophysical scene with the number of frames equal to the number of integrations.

\subsubsection{Components of Wavefront Error} \label{subsec:wfe}
The final observed PSF is a function of the wavefront error of the entire optical system (OTE + SI), geometric distortion, and detector-level effects such as plate scale and pixel response, all of which depend on field position to varying degrees. The primary mirror of JWST is made of 18 hexagonal mirror segments which are all aligned and co-phased to produce a single PSF \cite{Dean2006SPIE.6265E..11D}.  Each segment has mechanical supports with actuators which can be controlled to ensure the fidelity of the PSF.  Below, we summarize all components of the total OPD used to construct a simulated PSF \cite{Perrin2018SPIE10698E..09P, Rigby2023PASP..135d8001R}. As our technique depends on the estimation of the WFE, it is important to understand the various contributions and how those can be taken into account when producing synthetic PSFs.

\begin{figure} [ht]
\begin{center}
\begin{tabular}{c} %% tabular useful for creating an array of images 
\includegraphics[height=10cm]{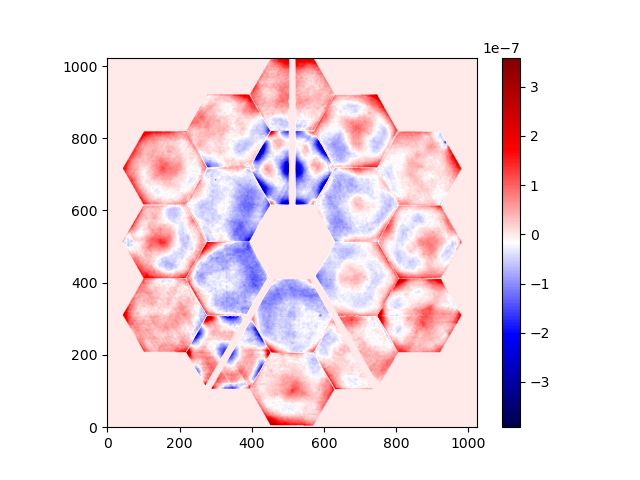}
\end{tabular}
\end{center}
\caption[static]
%>>>> use \label inside caption to get Fig. number with \ref{}
{ \label{fig:static} 
An example realization of high frequency wavefront errors of the JWST OTE. Each mirror segment was measured in lab individually and then combined with a zero gravity model to predict performance.  Units are meters. X and y units are pixels. High frequency errors on the OTE are static (unless impacted by events like micro-meteorite collisions) while low frequency errors (e.g. tip/tilt, defocus) can change for individual segments.}
\end{figure}

Some components of the overall WFE are time-invariant.  The OTE has static features caused by its physical characteristics, e.g. small imperfections due to mirror polishing, that are seen as high frequency errors.  In Fig. \ref{fig:static}, we show an example realization of the static wavefront error of the OTE, measured at the X-ray and Cryogenic Facility at NASA’s Marshall Space Flight Center \cite{McElwain2023PASP..135e8001M}. These measurements were conducted for each mirror segment individually and then combined with a zero gravity model to predict performance.

There are also low frequency errors from internal SI WFE that vary across the focal plane. These can be modeled by a set of Zernike polynomial coefficients, some examples of which are measured from the ISIM CV3 instrument test campaign and publicly available through WebbPSF for multiple field points across an instrument's focal plane \cite{Aronstein2016SPIE.9904E..09A}. However, these effects can be measured and are expected to be time invariant, allowing for their detailed modeling on sky given the position of a target \cite{Nardiello2022MNRAS.517..484N}. These static WFE variations arise from differences in how different portions of the internal optics are illuminated depending on the position of a star in the field. Ground cryovac testing campaigns have shown the SI WFE variations to be stable \cite{Sullivan2016SPIE.9951E..0ES}. A display of the the internal WFE for the NIRISS instrument at various positions along the field of view can be found in the WebbPSF Documentation \footnote{\url{https://webbpsf.readthedocs.io/en/latest/jwst.html\#id10}} \cite{Perrin2014SPIE.9143E..3XP}. There are also small OTE field-dependent WFEs, but they similarly introduce low frequency errors and their inclusion is not required to test our ability to recover the WFE of an observation thus not included in our model (see Sec. \ref{subsec:estimation}).

Lastly, micro-meteorite impacts can effectively change the time-invariant WFE of the OTE instantaneously by introducing new imperfections to the mirror segments \cite{McElwain2023PASP..135e8001M}.  These impacts will remain unaltered unless another micro-meteorite type event occurs in the same location.  Therefore, any micro-meteorite impacts must be included in any model of the WFE to accurately reproduce the real PSF of a given observation.  These are characterized during on-sky WFSC operations and can then be included in the OPD model.

\begin{figure} [ht]
\begin{center}
\begin{tabular}{c} %% tabular useful for creating an array of images 
\includegraphics[height=10cm]{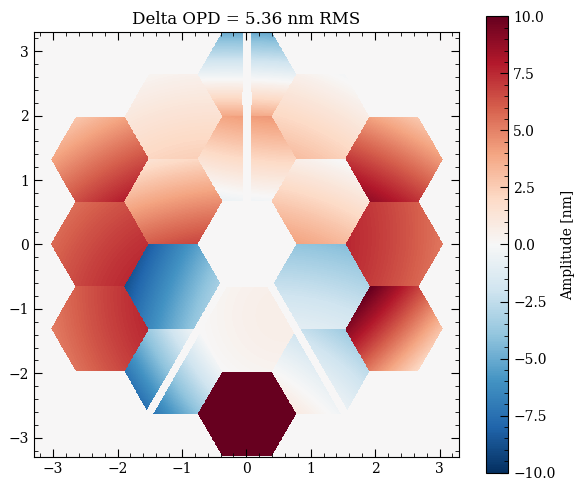}
\end{tabular}
\end{center}
\caption[opddrift]
%>>>> use \label inside caption to get Fig. number with \ref{}
{ \label{fig:opddrift} 
The difference in the OTE OPD after simulating some input amount of drift equivalent to rms residual WFE $\sim$ 5 nm.  Differences in the WFE are both specific to individual segments and affect the segments connected to each wing (leftmost and rightmost three segments).}
\end{figure}

In addition to static WFE features, there are dynamic contributions which alter the total wavefront error, which in turn affects the observed PSF. Two effects are due to changes to the angle of the sunshield relative to the incident solar radiation causing differential thermal changes across the primary mirror backplane. One effect is that the backplane will expand or contract.  Another effect is the thermal expansion of the frill which is connected to the primary mirror, generating motion of the segment supports and introducing segment specific wavefront drift.  A small component of the dynamic wavefront drift is due to oscillations from heater switching on the Integrated Science Instrument Module Electronics Compartment (IEC).  The last potential dynamic contribution to the wavefront error is a ``tilt event''.  These are large motions of individual mirror segments that typically occur only one segment at a time, seemingly due to changes of stress on the backplane, which can cause large changes in the wavefront error very quickly (adding $\sim$ 20-40 nm rms WFE) \cite{McElwain2023PASP..135e8001M}.  In the case of tilt events, they are discovered during WFS measurements and then corrected, where that WFS OPD can also be used to build an accurate model of the observed PSF. All of these components combined will change the total wavefront error between WFSC operations and the observation in question, resulting in a modified PSF.

These dynamic contributions will cause the WFE to drift over time, and therefore these factors must be included in modeling an example WFE for a given observation.  Leisenring et al. (in prep.) developed simulation tools (pyNRC) for various observing modes on NIRCam \footnote{\url{https://pynrc.readthedocs.io/en/latest/index.html}}.  Within pyNRC, they use webbpsf to simulate the processes listed above that can cause the OPD of the OTE to change over time.  These processes follow specific functional forms which will drift the OPD by an expected amount and with an expected pattern.  Given an input amount of time and pitch angle relative to the Sun, the effects from changes in the equilibrium temperature can be calculated and the difference in the WFE estimated.  As shown in Fig. \ref{fig:opddrift}, the wavefront drift is confined to individual segments and the wings of the primary mirror.  It follows a low frequency pattern across individual segments. The frill and IEC components are defined through the piston, tip, and tilt terms for each mirror segment. The thermal slew contribution is defined through a set of nine Hexikes for each mirror segment, and dominated by piston, tip, and tilt. In order to simulate the appropriate data for our technique, we first construct the OPD map (containing contributions from the OTE and SI) to serve as the WFSC measured OPD.  Then, we drift this OPD using pyNRC to represent a large motion of 45 degrees relative to the pitch angle of the Sun over the course of 24 hours, the expected longest time between WFSC observations.  That drifted OPD is used to generate a PSF with POPPY which is then imported into ami\_sim to include detector-level noise sources and generate a data cube of the target observation.

\subsubsection{Simulating a Real Observing Program} \label{subsec:realsim}

For our purposes, we chose to simulate an observing program that used NIRISS AMI for high-contrast imaging.  GO-1843 (PI: Kammerer) observed HD 206893 in order to directly image HD 206893 B in the F380M, F430M, and F480M filters.  HD 206893 B \cite{Milli2017A&A...597L...2M} has a dynamical mass estimate of 27.4 $^{+2.3}_{-2.3}$ M$_{Jup}$ \cite{Hinkley2023A&A...671L...5H} with a semi-major axis of 9.7$^{+0.4}_{-0.3}$ au from the star at 40.707 $\pm$ 0.067 pc \cite{Gaia2018A&A...616A...1G} and a very red color with a contrast of 7.91 mag in L' (3.8 $\mu$m) on VLT/NACO.  To generate a simulation of this target observation in AMI and FP modes, we extract the electron count rate (here 2.15 * 10$^{6}$ e$^{-}$/s in the AMI case and 1.138 * 10$^{7}$ e$^{-}$/s in the FP case) as input to \texttt{ami$\_$sim} from the JWST ETC using the planned observing setup in the F430M filter and scaling the flux to 2MASS K$_{s}$ = 5.593 mag with a spectrum of a blackbody with T$_{eff}$ = 6500 K for this F5V star.

The exposure configuration for GO 1843 consists of observations with the F430M filter using 2177 integrations of 8 groups for a total exposure time of $\sim$ 1523s.  \texttt{ami$\_$sim} produces the same number of frames in the data cube as integrations.  In order to avoid making a computationally prohibitive algorithm fitting to 2177 frames, we instead produce 14 frames to which we can compare our model.  The total exposure time is $\sim$ 25 min with each frame representing a co-added image over $\sim$ 1.8 min.  These 14 frames serve as co-added slices of the observation as the wavefront is not expected to change in any meaningful way during the course of this relatively short observation.  We conserve the number of total groups for the planned observation to correctly simulate the noise and the expected signal-to-noise.  Instead of 2177 integrations of 8 groups each, we produce a data cube of 14 integrations of 1244 groups in \texttt{ami$\_$sim} which does not include saturation effects.  We simulate the two observing modes for a target without a companion to define sensitivity limits (see Sec. \ref{subsec:limits}) and with a companion that has the expected features of HD 206893 B at the time of observation to determine our ability to recover and estimate the parameters of a known injected companion (see Sec. \ref{subsec:recovery}).  The target observation both with and without a companion was simulated using the drifted OPD to represent the expected rms residual WFE between the WFSC operations and the real observation, as stated in Sec. \ref{subsec:wfe}.

\subsection{Estimation} \label{subsec:estimation}

\subsubsection{Modeling the Wavefront Error} \label{subsec:modelwfe}

Our synthetic PSF model is generated from OPD maps using POPPY in the same manner as the simulations. Importantly, in applications to real data, polychromatic PSF models consistent with the known spectra of the target star should be used instead of the monochromatic PSF models we have used here for the simulations and modeling. Using only monochromatic PSF models on real data would result in residuals on the order of 0.1\% per pixel with a blurring effect in the peaks of the PSF and limit our ability to detect high contrast companions. Our simulations and models both use monochromatic PSF models for consistency. High frequency errors are static, and the portion of the WFE that is dynamic is either confined to the wings or to individual mirror segments, see Fig. \ref{fig:opddrift}.  Therefore, we chose the hexagonal mirror segment specific Zernike coefficients (Hexikes) to serve as the basis with which we construct an OPD map.  We use three Hexike terms per segment (piston, tip, and tilt) to realistically model the WFE without being computationally prohibitive as the WFE drift is limited to the piston, tip, and tilt coefficients (see Fig. \ref{fig:opddrift}).  Our model to reproduce full pupil images contains 54 Hexike parameters (3 for each of the 18 segments), and that for AMI images contains 21 Hexike parameters (3 for each of the 7 segments).

While Hexike coefficients are appropriate to model low frequency errors across individual mirror segments, high frequency errors cannot be well modeled with a reasonable number of Hexike coefficients.  Importantly, the high frequency errors are a direct result of mirror imperfections that are well measured on the ground and are continuously monitored through WFS observations. We can use the known high frequency errors to inform our model of the remaining segment specific low frequency WFE drift.  Our model of the total WFE of a given observation is a combination of the high frequency WFE (example shown in Fig. \ref{fig:static}) and the estimated low frequency error on a segment by segment basis where the Hexikes are free parameters.

\subsubsection{WFE Optimization and Companion Detection} \label{subsec:bestfit}
We perform a Bayesian analysis using the Nested Sampling Monte Carlo routine to derive the best fit PSF model to our simulations with the python module PyMultiNest, a fast sampling technique that derives the Bayesian evidence of the input model \cite{Feroz2009MNRAS.398.1601F, Buchner2014A&A...564A.125B, Golomb2021AJ....162..304G}.  PyMultiNest generates a user defined number of live points that sample the prior distribution and evaluates the likelihood of the sampled model solution.  It then lists these likelihood values in order from highest to lowest likelihood, and re-samples the prior distribution.  Samples with likelihoods higher than the previous list of likelihoods are then saved to that list with the rest rejected, ensuring a maximization of the likelihood of the model.  PyMultiNest continues this process of sampling and rejecting model solutions until the difference in the likelihood of new samples is negligible. The evidence is the integral of the likelihood times the prior over all parameters which allows us to test different models and determine which model is preferred given the data, e.g. if a star with a companion is preferred over just a star.

\begin{figure} [ht]
\begin{center}
\begin{tabular}{c} %% tabular useful for creating an array of images 
\includegraphics[height=5.3cm]{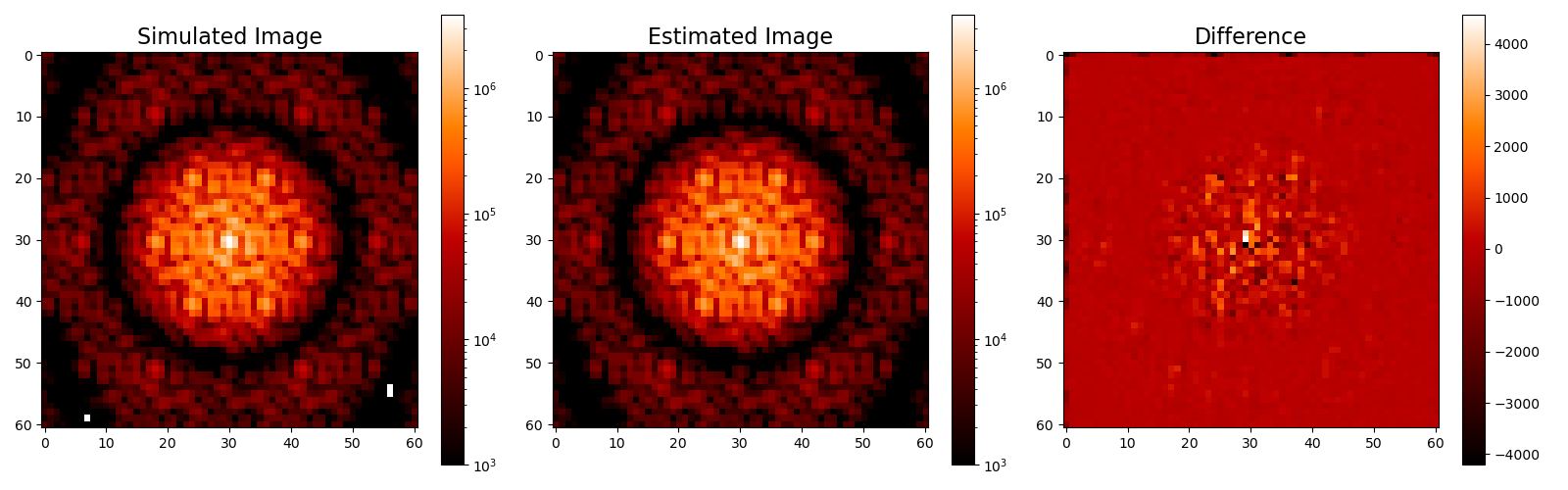}
\end{tabular}
\end{center}
\caption[amiimage]
%>>>> use \label inside caption to get Fig. number with \ref{}
{ \label{fig:amiimage} 
Left: First co-added frame from simulated data cube produced by ami\_sim in the AMI observing mode for NIRISS. This frame is representative of 155 co-added images each with eight groups. Center: Image produced with the maximum likelihood estimate of the WFE for the simulated image. Right: Residuals of first simulated frame minus our maximum likelihood estimated image. Note: Colorbar scales are the same for the left and center image, but different for the right image.}
\end{figure} 

\begin{figure} [ht]
\begin{center}
\begin{tabular}{c} %% tabular useful for creating an array of images 
\includegraphics[height=5.3cm]{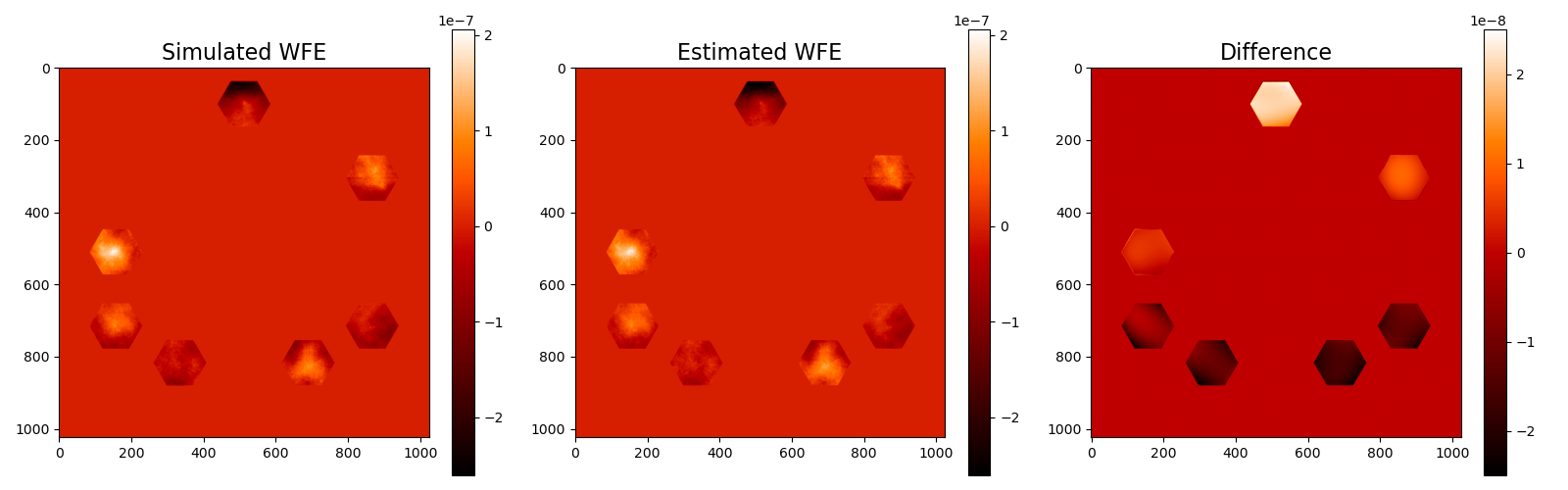}
\end{tabular}
\end{center}
\caption[amiwferesult]
%>>>> use \label inside caption to get Fig. number with \ref{}
{ \label{fig:amiwferesult} 
Left: Simulated WFE of target for AMI observation. Center: Maximum likelihood estimate of the WFE. Right: Residuals of simulation minus estimate. Note: Colorbar scales are the same for the left and center image, but different for the right image.}
\end{figure} 

\begin{figure} [ht]
\begin{center}
\begin{tabular}{c} %% tabular useful for creating an array of images 
\includegraphics[height=5.3cm]{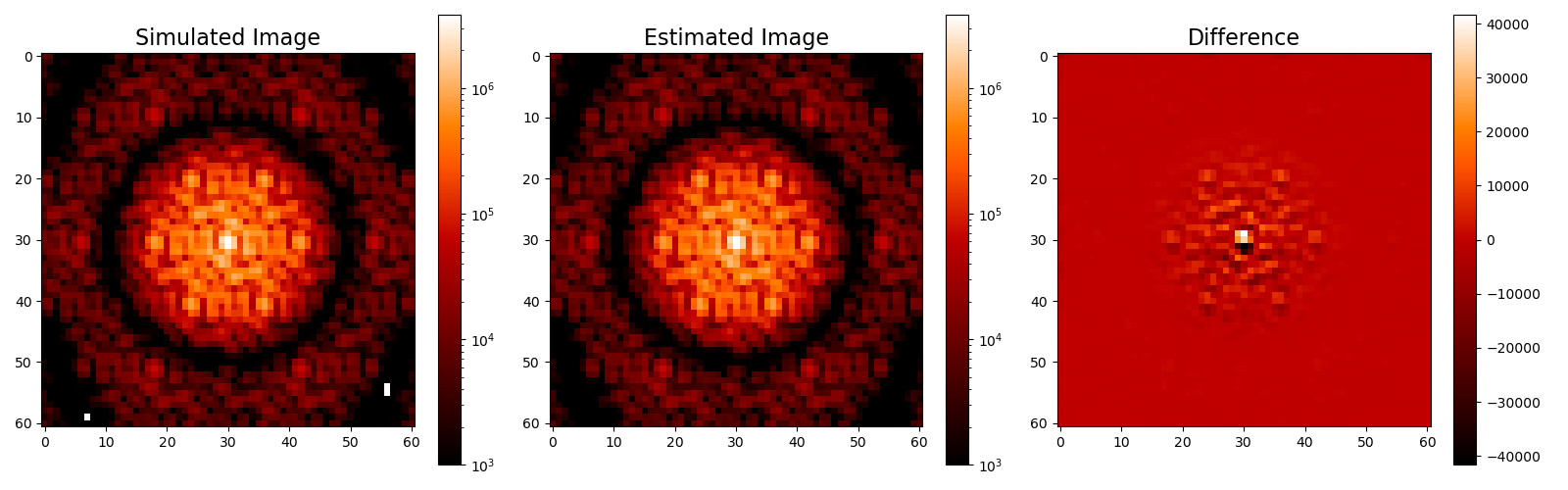}
\end{tabular}
\end{center}
\caption[amiwfscresult]
%>>>> use \label inside caption to get Fig. number with \ref{}
{ \label{fig:amiwfscresult} 
Left: First co-added frame from simulated data cube produced by ami\_sim in the AMI observing mode for NIRISS. This frame is representative of 155 co-added images each with eight groups. Center: Image produced from the WFSC OPD that was used to define the priors in our technique. Right: Residuals of first simulated frame minus image produced from WFSC OPD. Note: Colorbar scales are the same for the left and center image, but different for the right image. Residuals are higher compared to Fig. \ref{fig:amiimage}.}
\end{figure} 

Our free parameters are the three Hexike coefficients to each mirror segment (which combined with the static WFE produces a total WFE), three parameters describing the primary star (x and y position of the center of the PSF and flux), and three additional parameters for each additional source in the field (separation between center of primary star and center of companion, position angle of companion, and difference in magnitude between primary and companion). The total WFE is used to generate a PSF model from poppy, and the astrophysical scene is constructed using that PSF and the positions of all sources included in the model. We then evaluate the log-likelihood of the model by comparing the constructed astrophysical scene to all the frames of the simulation from \texttt{ami$\_$sim}. The log-likelihood is defined by a chi squared where n is the total number of pixels in all simulated data frames and $\sigma_{\gamma,i}$ is the photon noise in each pixel ($\sqrt(data_{i}$)):

\begin{equation}
\mathcal{L} =  \sum_{i=1}^{n} -0.5*(\frac{model_{i} - data_{i}}{\sigma_{\gamma,i}})^{2}
    \label{likelihood}
\end{equation}

For a scene with a single star, we center the data in the brightest pixel and set the priors as flat for the central position of the star: -0.13'' $\leq$ x$_{cen}$ $\leq$ 0.13'' and -0.13'' $\leq$ y$_{cen}$ $\leq$ 0.13''.  The flux of the star is defined as a normalization constant to the PSF model from POPPY where we define a flat prior between 10$^{5}$-10$^{12}$.  For a scene with a star and a companion, we define the stellar priors the same as above, and define the companion priors as flat: 0.5 pixels $\leq$ separation $\leq$ 10.0 pixels from the center, 0 degrees $\leq$ position angle $<$ 360 degrees, and 3.0 mag $\leq$ $\Delta$mag  $\leq$ 10.0 mag.

Importantly, we must also set the priors to each Hexike coefficient that will define the estimated WFE of the target observation. As stated in Sec. \ref{subsec:wfe}, a realistic observation for our high-contrast imaging program includes a realization of an OPD map intended as a stand-in for the measured OPD during WFSC operations that is then drifted to produce the target observation. To set the priors for the Hexike coefficients in estimating the WFE of the target observation, we take this WFSC OPD map, subtract the static WFE, and deconstruct the residual into its constituent Hexike coefficients for each mirror segment using the decompose\_opd function in POPPY and a mask for each segment.  We also perform the same task for the drifted OPD used to represent the WFE at the time of observation to determine the amount any individual Hexike is expected to drift.  The maximum difference in any Hexike between the WFSC OPD and the drifted OPD then serves as the amplitude for the priors of each Hexike, centered around the Hexike value of the deconstructed WFSC OPD.  We found that at most, a given Hexike term will drift 20 nm, and therefore our priors are the deconstructed WFSC OPD Hexike terms $\pm$ 25 nm.

\section{Results} \label{sec:results}

We applied our companion detection and WFE estimation technique on the simulations of the observing program GO 1843 to determine how well we expect to recover companions.  We performed this analysis on both AMI and FP simulations of single stars and a star with an injected companion. Then, we compared our expected performance to other post-processing techniques that utilize the observation of a reference star.

\subsection{Sensitivity Limits of Technique} \label{subsec:limits}
To derive sensitivity limits, we applied our WFE estimation technique to a simulation of an observation of a single star, defined in Sec. \ref{subsec:realsim}.

In the AMI case, we show the image produced from the maximum likelihood estimate of our 24 parameter model (3 Hexikes for each of the 7 mirror segments, 3 parameters for the position and flux of the star) relative to the first frame within the simulated data cube in Fig. \ref{fig:amiimage}.  In Fig. \ref{fig:amiwferesult}, we show our ability to recover the WFE of the simulated data.  The pattern of the residual WFE corresponds to higher order Zernike polynomials, which were not included in our model. Higher order terms contribute a small amount to the overall WFE, but their inclusion as free parameters to the model introduces degeneracies that prevent accurate estimates of the piston, tip, and tilt terms of each mirror segment within a reasonable run-time of the algorithm. In this case, the code often over-fits the data and the dominant first three coefficients are less well constrained, hence our use of 3 Hexikes per mirror segment. Importantly, the wavefront drift of higher order terms due to thermal effects is a function of the physical piston, tip, and tilt drift of a given segment. Each segment is curved and any change in the piston, tip, and tilt terms will necessarily cause changes to the higher order terms, but in a way that is physically well-defined. In the example estimation of simulated data, we did not estimate the higher order terms of each mirror segment from the original WFSC simulation or thermal drift of piston, tip, and tilt. With well estimated higher order terms from the WFSC data and the known motion from specific thermal drift, we can better model the final WFE of the simulated data. A future iteration of our technique will include the definition of these higher order terms from the estimated piston, tip, and tilt free parameters. In Fig. \ref{fig:amiwfscresult}, we also show the image produced from subtracting the PSF generated from WFSC OPD from the simulated images.  In comparing the residuals in Figs. \ref{fig:amiimage} and \ref{fig:amiwfscresult}, the WFE drift introduces features in the fringes of the AMI PSF, both in the center and where the sub-apertures constructively interfere to create additional peaks in the image further from the center. Our WFE estimation technique models the change in OPD that creates these features allowing for a deeper search for companions.

\begin{figure} [ht]
\begin{center}
\begin{tabular}{c} %% tabular useful for creating an array of images 
\includegraphics[height=5.3cm]{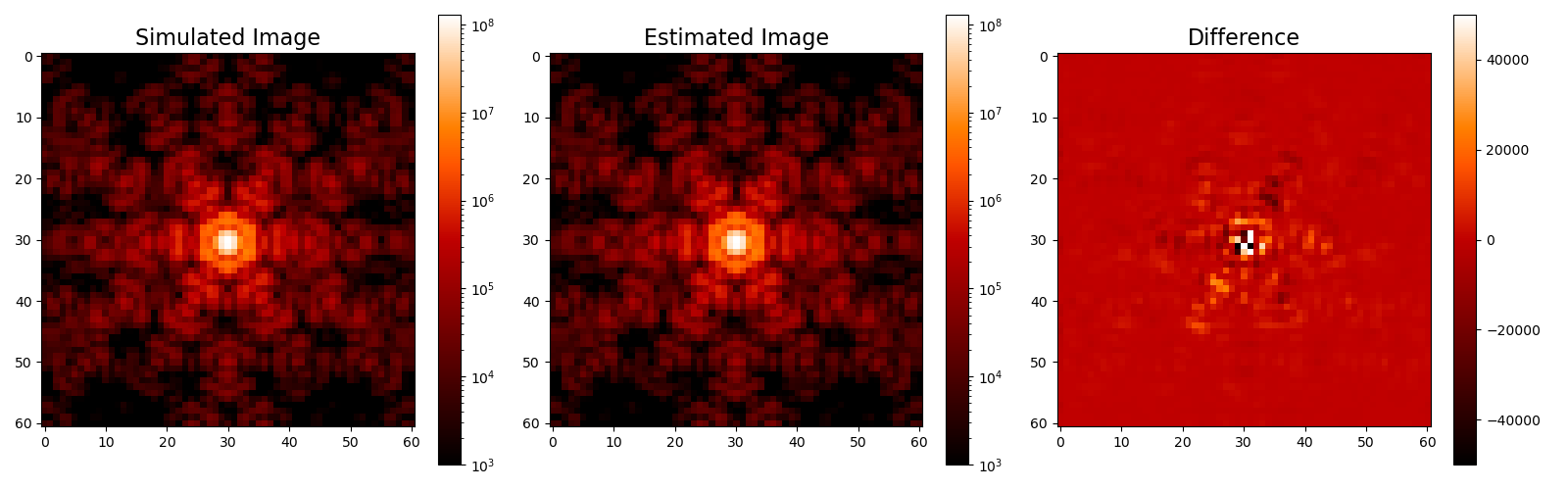}
\end{tabular}
\end{center}
\caption[fpimage]
%>>>> use \label inside caption to get Fig. number with \ref{}
{ \label{fig:fpimage} 
Left: First co-added frame from simulated data cube produced by ami\_sim in the full pupil observing mode for NIRISS. This frame is representative of 155 co-added images each with eight groups. Center: Image produced with the maximum likelihood estimate of the WFE for the simulated image. Right: Residuals of first simulated frame minus our maximum likelihood estimated image. Note: Colorbar scales are the same for the left and center image, but different for the right image.}
\end{figure}

\begin{figure} [ht]
\begin{center}
\begin{tabular}{c} %% tabular useful for creating an array of images 
\includegraphics[height=5.3cm]{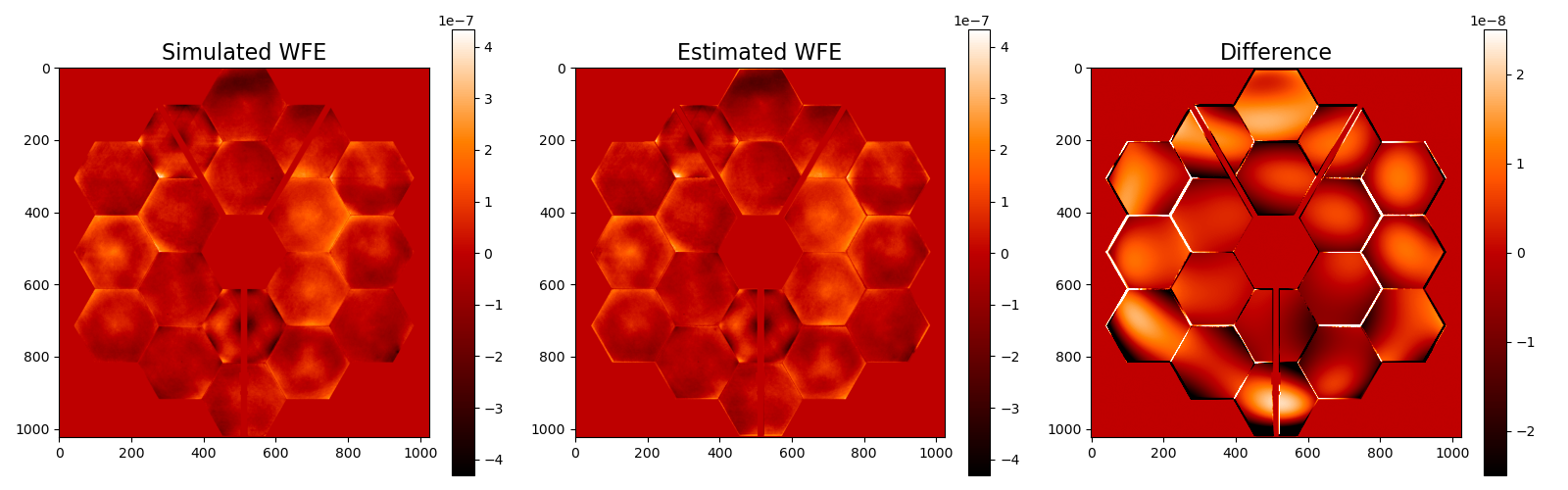}
\end{tabular}
\end{center}
\caption[fpwferesult]
%>>>> use \label inside caption to get Fig. number with \ref{}
{ \label{fig:fpwferesult} 
Left: Simulated WFE of target for full pupil observation. Center: Maximum likelihood estimate of the WFE. Right: Residuals of simulation minus estimate. Note: Colorbar scales are the same for the left and center image, but different for the right image.}
\end{figure}

\begin{figure} [ht]
\begin{center}
\begin{tabular}{c} %% tabular useful for creating an array of images 
\includegraphics[height=5.3cm]{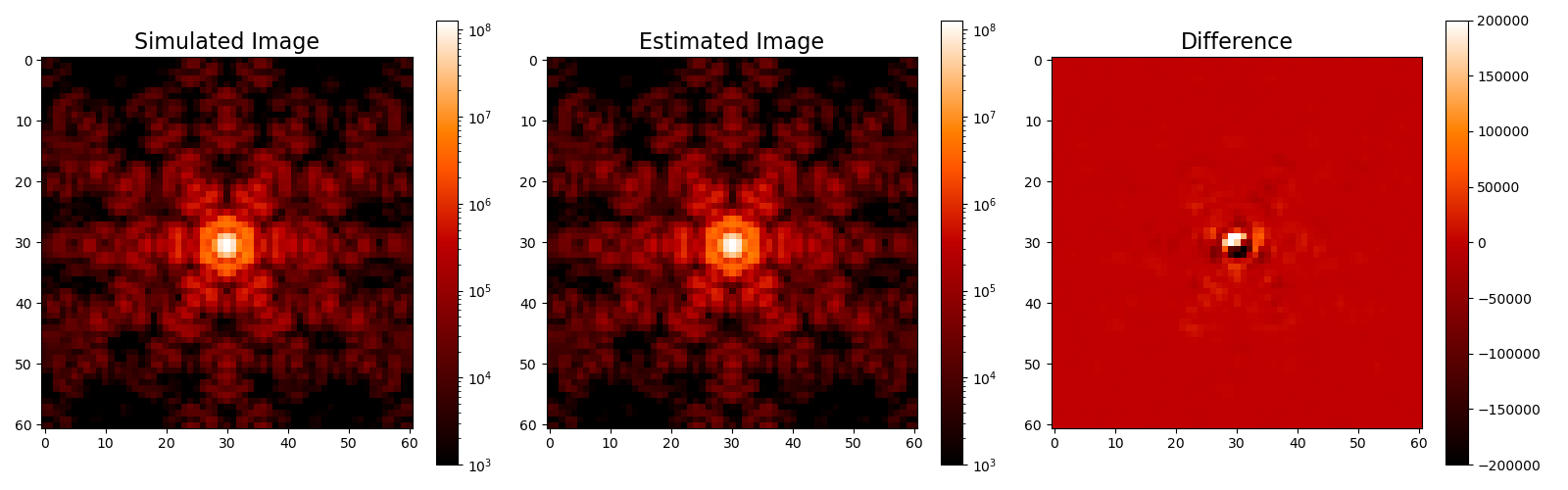}
\end{tabular}
\end{center}
\caption[fpwfscresult]
%>>>> use \label inside caption to get Fig. number with \ref{}
{ \label{fig:fpwfscresult} 
Left: First co-added frame from simulated data cube produced by ami\_sim in the full pupil observing mode for NIRISS. This frame is representative of 155 co-added images each with eight groups. Center: Image produced from the WFSC OPD that was used to define the priors in our technique. Right: Residuals of first simulated frame minus image produced from WFSC OPD. Note: Colorbar scales are the same for the left and center image, but different for the right image. Residuals are higher compared to Fig. \ref{fig:fpimage}.}
\end{figure}

We also applied our technique to simulations of the full pupil observing mode.  For this mode, we show the image constructed from the maximum likelihood estimate of our model compared to the simulated data in Fig. \ref{fig:fpimage}.  In Fig. \ref{fig:fpwferesult}, we show the recovered WFE relative to the WFE used to simulate the target observation. As with the AMI case, the residual WFE appears to follow the pattern associated with higher order Zernike polynomials. The estimate of the wavefront with 54 parameters already requires long computation time, and any increase in the number of coefficients per segment as free parameters would both increase degeneracy and make computation times prohibitively long. As mentioned above, a future iteration of this technique will include the known physical dependence of higher order terms on piston, tip, and tilt drifts as well as an estimation from the WFSC data to improve the model. In Fig. \ref{fig:fpwfscresult}, we also show the image produced from subtracting the PSF generated from WFSC OPD from the simulated images. In comparing the residuals in Figs. \ref{fig:fpimage} and \ref{fig:fpwfscresult}, the WFE drift introduces large features in the PSF at separations close to the star. Our WFE estimation technique models the change in OPD that creates these features allowing for an improved search for companions at close separations.

\begin{figure} [ht]
\begin{center}
\begin{tabular}{c} %% tabular useful for creating an array of images 
\includegraphics[height=6.35cm]{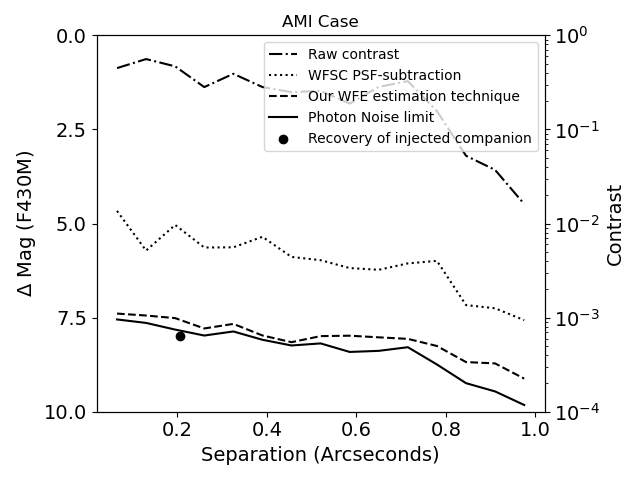}
\includegraphics[height=6.35cm]{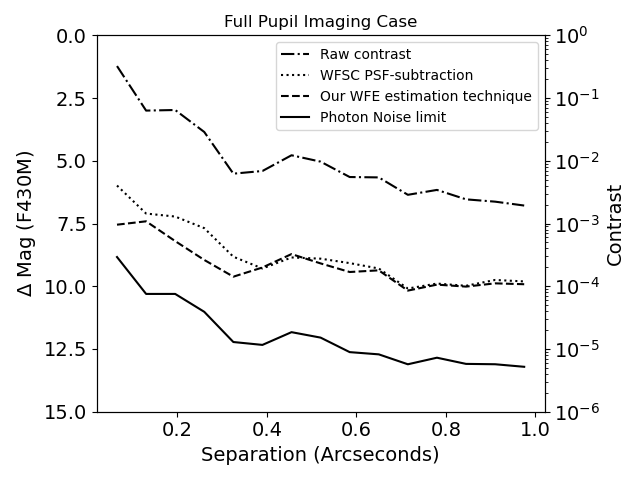}
\end{tabular}
\end{center}
\caption[amicc]
%>>>> use \label inside caption to get Fig. number with \ref{}
{ \label{fig:amicc} 
5$\sigma$ contrast curve for the AMI case (left) and the FP case (right) for the simulation of a single star described in Sec. \ref{subsec:realsim}. The 5$\sigma$ result is calculated over the sub-apertures within successive annuli over all 14 images of the simulation as described in Sec. \ref{subsec:limits}. This simulation is similar to typical high contrast imaging programs. The dot-dashed line corresponds to the raw contrast, the dotted line corresponds to the contrast curve when using the WFSC OPD to reconstruct a single star, the dashed line corresponds to the contrast curve derived after the application of our WFE estimation method, and the solid line corresponds to the photon noise limit. Raw contrast and photon noise limits are different in these two cases as the full pupil case attains a higher S/N thus higher photon noise limit, and the AMI PSF is more spread out due to interference of the sub-apertures corresponding to reduced raw contrast over the field of view of the sub-array in question. The black dot represents the values of the source injected and recovered in Sec. \ref{subsec:recovery}. Our Bayesian wavefront estimation improves on current PSF subtraction methods that utilize the temporally closest WFSC measurements of the OPD over all separations in the AMI case and for $<$ 0.4'' in the FP case.}
\end{figure}

In Fig. \ref{fig:amicc}, we show the contrast curves derived from our maximum likelihood estimate of the WFE of the target observation for both the AMI and FP observing modes, respectively. We measure our ability to recover companions by taking the difference between each frame of the simulated observation and the image produced from the maximum likelihood estimate model. Then, we define successive annuli, centered on the star, with a 0.14'' width equivalent to the full width at half maximum (FWHM) in the F430M filter ($\sim2$ pixels). Within each annulus, we construct adjacent circular sub-apertures with diameters equivalent to the FWHM and calculate the flux within each sub-aperture over all frames.  We then evaluate the t-test distribution for a 5$\sigma$ detection given the calculated standard deviation and number of sub-apertures within the annuli. We use the approach of \cite{Mawet2014ApJ...792...97M} (in their Sec. 3.4) to correct for small sample sizes (i.e. at small angles due to a low number of sub-apertures) to arrive at the flux that corresponds to the 5$\sigma$ value above the residuals within each annulus. We calculate the flux of the primary star by summing the pixels within an aperture with diameter equivalent to the FWHM centered on the primary. With this measurement, we can compare the detectable flux to that of the primary for an achievable relative contrast measurement in flux ratio or $\Delta$mag. We also perform this calculation for the raw contrast of the simulated images, the contrast after subtracting the PSF generated from the WFSC OPD, and the contrast after subtracting the simulated images with no noise from those with only photon noise to define the photon noise limit. The difference between the WFSC OPD results and that of our method demonstrate the improvement our technique generates after WFE drift of the telescope.

\subsection{Traditional Interferometric Analysis with Calibrator} \label{subsec:comparison}

Interferometric analysis is commonly performed on AMI imaging data in order to search for faint, close companions. Interferometric analysis techniques on AMI data take advantage of a non-redundant aperture mask, such as that utilized on NIRISS \cite{Greenbaum2015ApJ...798...68G, Sivaramakrishnan2023PASP..135a5003S} where the squared visibilities and closure phases are calculated from the data. Closure phases are an interferometric quantity derived from the fringes of each pair of apertures within a closed triangle of three baselines (e.g. three sub-apertures), utilized to remove phase errors  \cite{Jennison1958MNRAS.118..276J,Rogers1974ApJ...193..293R, Monnier2000plbs.conf..203M, Monnier2004ApJ...602L..57M}. Any scene deviating from a point-symmetric object (e.g. a binary or an extended structure) will produce non-zero closure phases with amplitudes on the order of the amount of deviation (e.g. flux ratio). This technique relies on the observation of a reference star in order to account for any bias in the closure phases as a result of wavefront drift, thus increasing total program observing time.

This technique has been applied to simulations of JWST data \cite{Sallum2019JATIS...5a8001S}. We apply their same algorithm to the simulation produced in our paper to determine the sensitivity limits of this technique. We use a single star as a reference that was produced from an OPD with rms residual WFE of 5 nm relative to the simulated target observation (roughly the expected WFE drift through the course of a full JWST reference-target observation). This serves as the calibrator source to remove bias, after which the derived closure phases and squared visibilities are used to estimate sensitivity.

\begin{figure} [ht]
\begin{center}
\begin{tabular}{c} %% tabular useful for creating an array of images 
\includegraphics[height=9cm]{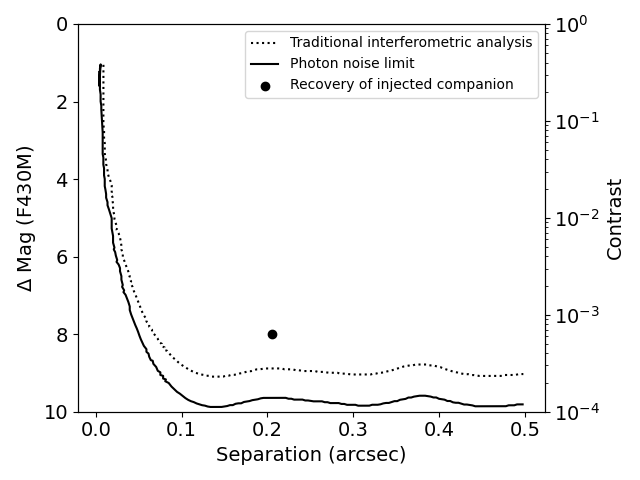}
\end{tabular}
\end{center}
\caption[amisensitivitysteph]
%>>>> use \label inside caption to get Fig. number with \ref{}
{ \label{fig:amisensitivitysteph} 
Contrast curve for the AMI case using traditional interferometric post-processing analysis \cite{Sallum2019JATIS...5a8001S} on the observables, to compare to our WFE estimation technique shown in Fig. \ref{fig:amicc}. The dotted line is the sensitivity limit that corresponds to a $\chi^{2}$ - $\chi^{2}_{single}$ = 25, where $\chi^{2}$ is the value associated with a single companion model. The solid line is the theoretical photon noise limit for these data derived with the technique of \cite{Ireland2013MNRAS.433.1718I}. This contrast curve is extended to small, sub-pixel separations as the sensitivity is defined based on the $\Delta\chi^{2}$ calculation from closure phases.}
\end{figure}

Because these observables are in the Fourier plane and not the image plane, the definition of the sensitivity to companions cannot be the same as done in Sec. \ref{subsec:limits}. In this case, the closure phases and squared visibilities of the simulated single star were extracted and then fitted with a grid of single companion models with varying separation, position angle, and contrast. For each grid point, the difference in $\chi^{2}$ is calculated between that binary model and the null (no companion) model. These values are averaged over position angle, and for each value in separation a $\Delta\chi^{2}$ interval is then used to determine the single companion contrast that would be distinguishable from the no companion model with 5$\sigma$ significance. This detection limit is defined as $\chi^{2}$ - $\chi^{2}_{single}$ = 25. We also estimate the theoretical photon noise limit for these data with the interferometric observables using the method of \cite{Ireland2013MNRAS.433.1718I}. These limits are shown in Fig. \ref{fig:amisensitivitysteph}.

\begin{figure} [ht]
\begin{center}
\begin{tabular}{c} %% tabular useful for creating an array of images 
\includegraphics[height=9cm]{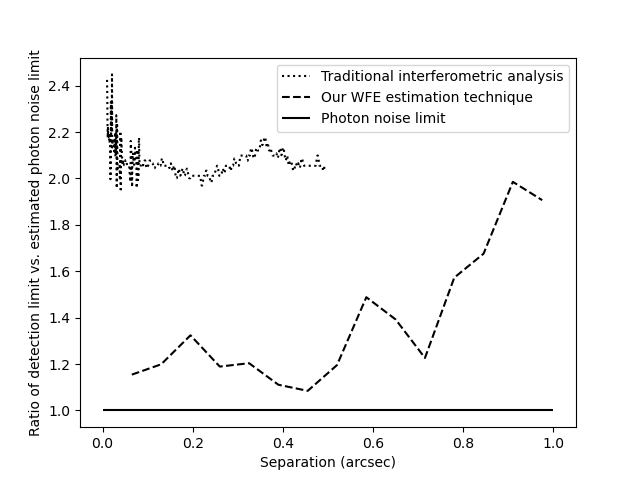}
\end{tabular}
\end{center}
\caption[cccompare]
%>>>> use \label inside caption to get Fig. number with \ref{}
{ \label{fig:cccompare} 
Comparison of the sensitivity limit of each technique in the AMI case to its estimated photon noise limit. The two techniques must be evaluated in different ways, and this comparison shows the ability relative to its theoretical limit estimates. Both techniques approach their theoretically derived photon noise limits.}
\end{figure} 

The sensitivity of traditional interferometric analysis is estimated using the interferometric observables (not the images). The sensitivity of our technique is estimated using the residuals in the image plane.  Therefore, we cannot use the same statistical method to compare techniques. We can compare the ability of each technique relative to its calculated theoretical detection limit, the photon noise limit, which is estimated in the respectively plane of each technique. These limits are arrived at in differing ways, but give an indication as to how these methods compare. Shown in Fig. \ref{fig:cccompare} is the ratio of the 5$\sigma$ recovery limit to the photon noise limit for each method. The traditional interferometric analysis is sensitive at roughly twice the photon noise limit (due to the 5 nm rms residual WFE from the reference star to the target), while our technique is sensitive between roughly 1.1-2.0x the photon noise limit across 1''. The absolute attainable contrast with the traditional interferometric analysis is roughly 9 mag beyond 0.1'', while ours is roughly 7.5 mag, but both approach the estimated photon noise limit of each technique since they are measured in different ways.

\subsection{Companion Recovery} \label{subsec:recovery}
Both our WFE estimation technique and the technique of \cite{Sallum2019JATIS...5a8001S} are designed to detect and characterize faint companions. As a more direct test of the ability of each technique, we created a simulation of a faint companion and attempted to recover the signal with both methods. We simulated the expected scene for GO 1843 with a companion at 0.2055'' in separation from the central star ($\sim$ 1.5 $\lambda/D$), $\Delta$mag = 8 mag in the F430M filter, and a position angle of -71.57$^{\circ}$.

\begin{figure} [ht]
\begin{center}
\begin{tabular}{c} %% tabular useful for creating an array of images 
\includegraphics[height=8cm]{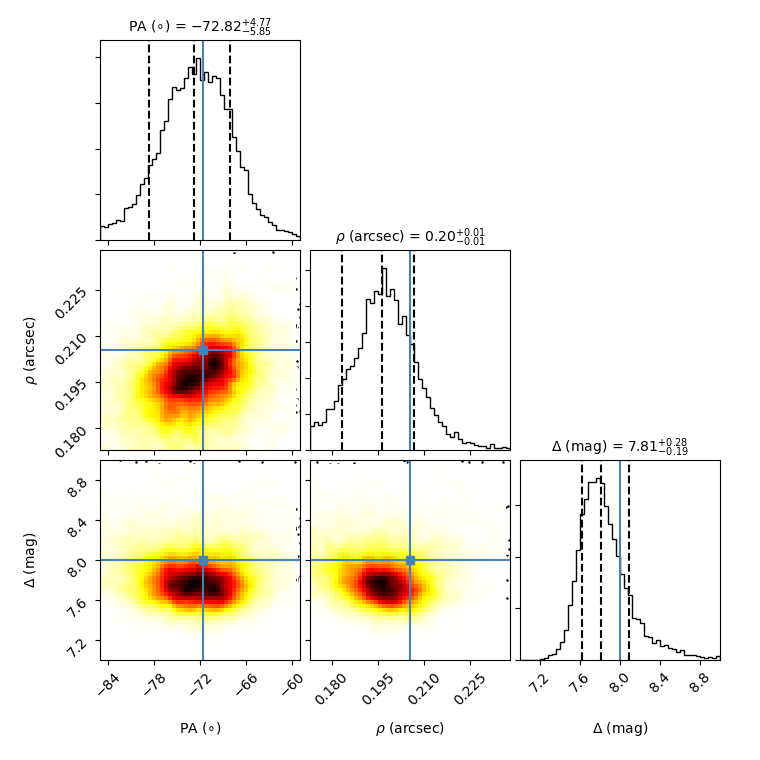}
\end{tabular}
\end{center}
\caption[amicompanion]
%>>>> use \label inside caption to get Fig. number with \ref{}
{ \label{fig:amicompanion} 
Posterior distributions for the companion parameters derived for HD 206893 B using our novel technique. Blue lines are the simulated values, and left/right black dashed lines enclose 68\% of the marginalized posterior around the median (central black dashed line). Displayed values are the median with the enclosed 68\% of the marginalized posterior.}
\end{figure}

We applied our WFE estimation and companion search method on these data. Shown in Fig. \ref{fig:amicompanion} are the posteriors of the companion parameters that were estimated with our technique.  The maximum likelihood estimate gives the following parameters: separation of the companion = 0.195'' , position angle = -70.7$^{\circ}$, and contrast = 7.85 mag. The median and the enclosed 68\% of the marginalized posterior are given in Table \ref{table1}. These estimated companion parameters are within the 68\% confidence interval encompassing the known companion parameters. In addition, we applied our WFE estimation routine to this data fitting only a single star and no companion. After evaluating the Bayesian evidence of the fit with and without a companion, we find a difference of 1300 in favor of the fit with a companion, evidence for a strong detection  \cite{Trotta2008ConPh..49...71T}. %See Fig. \ref{fig:wfepluscomp} for the posteriors of both the companion parameters and the Hexike coefficients of each mirror segment that define the WFE estimate. 

\begin{figure} [ht]
\begin{center}
\begin{tabular}{c} %% tabular useful for creating an array of images 
\includegraphics[height=8cm]{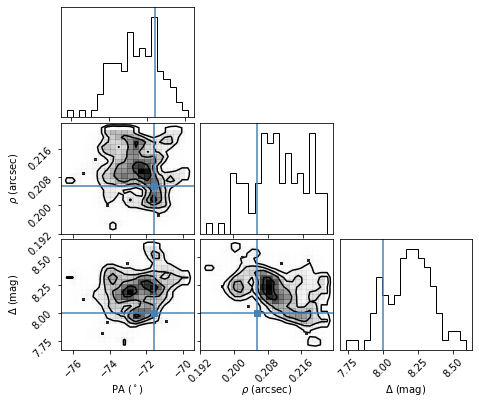}
\end{tabular}
\end{center}
\caption[amicompanionsteph]
%>>>> use \label inside caption to get Fig. number with \ref{}
{ \label{fig:amicompanionsteph} 
Posterior distributions for the companion parameters derived for HD 206893 B using traditional interferometric analysis techniques in the interferometric observables.  Blue lines are the true simulated values.}
\end{figure} 

\begin{figure} [ht]
\begin{center}
\begin{tabular}{c} %% tabular useful for creating an array of images 
\includegraphics[height=6cm]{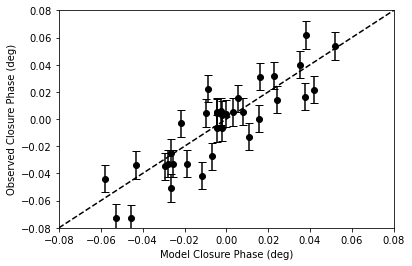}
\end{tabular}
\end{center}
\caption[closurephase]
%>>>> use \label inside caption to get Fig. number with \ref{}
{ \label{fig:closurephase} 
Closure phases of the model with a companion compared to the measured closure phases. This is indicative of a companion detection.}
\end{figure}

We then applied the traditional technique \cite{Sallum2019JATIS...5a8001S} to the same data and show the posteriors of the companion estimate in Fig. \ref{fig:amicompanionsteph}. We also show the measured closure phases of the simulated data and the fitted closure phases of the binary model in Fig. \ref{fig:closurephase} that are indicative of a binary detection. The real companion signal is being detected with this technique as well. The estimated companion parameters, displayed in Table \ref{table1}, are: PA($\circ$) = -72.53$^{+1.31}_{-1.20}$, separation ($\rho$)= 0.210$^{+0.006}_{-0.008}$, and contrast = 8.19$^{+0.18}_{-0.15}$ mag. The companion is detected within roughly the 68\% confidence interval for each parameter, and underestimates the contrast likely due to an underestimate of the closure phase errors. The error bars for the fitted companion parameters are comparable to those of our technique. It appears that in detecting such a faint companion, the ability to estimate accurate parameters are comparable between techniques.

\begin{table}[ht]
\caption{True parameters of the injected companion, estimate of the companion values with traditional interferometric analysis, and estimate of the companion values with our WFE estimation technique (median with 68\% marginalized posterior enclosed).}
\label{table1}
\begin{center}       
\begin{tabular}{|l|l|l|l|} 
\hline
\rule[-1ex]{0pt}{3.5ex}  & Separation & Position angle & $\Delta$mag  \\
\rule[-1ex]{0pt}{3.5ex}  & (arcsec) & (degrees) & (F430M)    \\
\hline
\rule[-1ex]{0pt}{3.5ex}  Truth	&	0.2055	&	-71.57	&	8.0	\\
\hline
\rule[-1ex]{0pt}{3.5ex} Traditional Interferometric	Analysis& 0.210$^{+0.006}_{-0.008}$	&	-72.53$^{+1.31}_{-1.20}$	&	8.19$^{+0.18}_{-0.15}$	\\
\hline
\rule[-1ex]{0pt}{3.5ex} Our WFE	Estimation Technique &	0.20$\pm$0.01	&	-72.82$^{+4.77}_{-5.85}$	&	7.81$^{+0.28}_{-0.19}$	\\
\hline 
\end{tabular}
\end{center}
\end{table}

Importantly, the injected companion has parameters that are below the 5$\sigma$ limit of our technique shown in Fig. \ref{fig:amicc}. Our estimate of the sensitivity limit is based on aperture photometry. However, the PSF of an AMI image has significant azimuthal structure and is well spread out, making our image plane WFE estimation technique more sensitive to faint companions than a 5$\sigma$ aperture photometry limit can illustrate. Therefore, this injection/recovery approach displays the power and similar performance of both the traditional interferometric technique and our WFE estimation technique. To attain more precise limits and comparisons, future work may depend on rigorous injection/recovery and false positive analysis \cite{DeFurio2022ApJ...925..112D}.

\section{Discussion} \label{sec:discussion}
\subsection{Utility of method} \label{subsec:abilityandutility}

Our technique utilizes WFSC information (scheduled to be taken every 48 hours) to inform a WFE estimation routine that models the OTE OPD for a given data set. This technique is paired with a companion search that allows for the simultaneous fitting of the WFE, primary star parameters, and companion parameters. With the success of this routine, we have proven the effectiveness of using synthetic PSF models to estimate the WFE of the target observation and perform high contrast imaging post-processing for full pupil and AMI data on the NIRISS instrument. The use of synthetic PSFs and WFSC information as priors to estimate the WFE reduces observing time by nearly 50\%, providing significant efficiency for highly coveted JWST time.

As shown in Figs. \ref{fig:amiwfscresult} and \ref{fig:fpwfscresult}, the WFE drift accumulated from changes in equilibrium temperature of the telescope will introduce significant differences in the PSF, especially at close separations. The application of our technique provides a marked improvement on the modeling of the PSF, allowing for the search for companions at close separations, see Figs. \ref{fig:amiimage} and \ref{fig:fpimage}. For the AMI case, the constructive interference of the light through the seven sub-apertures creates bright fringes far from the core of the PSF, distinct from the full pupil case.  This PSF is more sensitive to small changes in an individual Hexike coefficient. Combined with the small number of Hexike coefficients required to model the WFE, this results in a reduction in degeneracy between Hexike terms and allows for the improved modeling of the WFE and better achievable contrast, see Fig. \ref{fig:amicc}. In the FP case, the improvement from our WFE modeling is concentrated in the core of the PSF where we gain sensitivity over the WFSC model at separations $<$ 7 pixels (0.455'').

We also set out to determine if we could recover companions expected from scheduled JWST programs. We simulated a companion at 0.2055'' in separation and 8 magnitudes in contrast in the F430M filter, as expected for JWST observations of HD 206893 (GO-1843). We are able to recover the companion while simultaneously estimating the WFE, see Fig. \ref{fig:amicompanion}. As shown in Fig. \ref{fig:amicompanionsteph}, the companion parameter estimates from our novel technique in the AMI observing mode are comparable to that of the traditional approach \cite{Sallum2019JATIS...5a8001S} applied to the same simulated data. Although the estimated sensitivity limits differ between techniques, the approach to derive these limits are in two different planes. Therefore, the injection and recovery of a faint companion at a close separation shows the direct ability of both techniques, the traditional interferometric technique with a reference star and our WFE estimation technique without a reference star.

\subsection{Future Development and Applications} \label{subsec:futuredev}
There are several ways to improve on this technique and topics to explore using synthetic PSFs. In Fig. \ref{fig:amicc}, we show the contrast curves for this technique in AMI and FP modes, respectively. Full pupil PSF modeling is difficult due to the many degeneracies among the mirror segments and Hexike values. However, we can improve on WFE modeling if we simultaneously fit FP and AMI images \cite{Greenbaum2016OExpr..2415506G}. Our fitting routine converges on the Hexike values of each segment for AMI quickly, but it is a much slower process with just FP data.  If we fit both simultaneously, then the Hexikes for 7 of the 18 segments used in AMI observing will converge quickly, and that information will be able to break many degeneracies with the 11 other segments. As AMI imaging blocks $\sim$ 80\% of the light from the target, one would only need $\sim$ 20\% increased observing time to have comparable flux in both observing modes.

In this paper, we applied our novel technique to a single source, like will be performed in most high-contrast imaging programs. However, it should be noted that this type of technique can be applicable to wide-field images as well. The position along a detector will introduce low frequency WFE, as described in Sec. \ref{subsec:wfe}, due to internal optics and the position along the focal plane. This effect can be extensively modeled with efforts currently ongoing \cite{Nardiello2022MNRAS.517..484N}. The WFE contribution from the primary mirror segments will be constant, but other features of the optical system like the tertiary mirror introduce field dependent WFE specific to the instrument and position on the detector in question. With an estimate of low frequency errors based on position, we can also model the WFE from the OTE using all sources in the field of view as inputs. This type of analysis can also be useful when applied to future space missions like the Nancy Grace Roman Space Telescope (Roman) with the Wide-Field Instrument (WFI), a 0.28 square degrees field of view, where thousands of sources can be simultaneously observed to inform the WFE model. Due to the large field of view of the WFI, there will be large low-order field dependent WFEs which can be modeled and used to refine the WFE of the primary mirror.

Importantly, we showed that the residual WFE unaccounted for with our technique appears to be constructed of higher order Hexike coefficients. Because the drift is known to be confined to the first three Zernike coefficients of each mirror segment, the residual WFE from higher order coefficients is a result of not knowing the WFE associated with a given detector position. With the measured description of the low order WFE based on detector position, we can also improve upon our high contrast imaging approach as we model the three coefficients per segment that contain the drift.

Reference PSF modeling can also be useful for traditional interferometric post-processing techniques. The extracted interferometric observables are fit with models that describe the expected scene (e.g. one point source vs. two point sources).  Importantly, biases exist in these observables that must be precisely calibrated, hence the observation of a reference star directly before or after the target. However, it is possible to instead use the WFSC information to produce PSF models and calibrate the bias in the interferometric observables. This procedure would also increase the efficiency of any observing plan by removing the observation of the reference star, like with our technique.

This technique is planned to be applied as part of the Cycle 1 program, GO-2627 (PI: M. Ygouf).  In this program, JWST WFSC operations will be performed to derive the OTE OPD. Then, successive observations with NIRCam full pupil imaging and coronagraphic imaging are performed on a target, followed by another set of WFSC observations to again evaluate the OTE OPD which may have changed due to WFE drift from slewing. To determine the ability to estimate the WFE of a given observation, a technique similar to what we presented here will be applied to the full pupil and coronagraphic imaging. These estimates will be compared to the measured OTE OPD to determine how well this type of analysis performs on these different observing modes.

Other future space missions, like Roman, will also focus heavily on coronagraphic imaging with the goal of achieving contrasts of 10$^{-9}$. One feature of Roman is its active wavefront control which is intended to maintain a highly stable PSF during the course of observations \cite{Kasdin2020SPIE11443E..1UK}. The low order wavefront sensing and control is viable for V $<$ 5 mag and operates continuously to estimate the first 11 Zernike coefficients across the entire primary mirror. Higher order wavefront errors can be estimated during reference star observations, and corrected for during target observations. Additional higher order wavefront sensing measurements and commands will be evaluated and given from the ground that can inform post-processing analyses \cite{Kasdin2020SPIE11443E..1UK}.

Even given these processes to reduce WFE drift, a residual WFE will still remain due to WFE drift which can potentially be modeled by a technique like our own , although this is still speculative and requires substantial effort. Given that the Roman primary mirror is monolithic, it would require a model that uses a set of Zernike coefficients at high order (the degree to which remains to be seen \cite{Ygouf2020SPIE11443E..3NY}) instead of the Hexike approach with JWST. Additionally, the multiple pupil planes from coronagraphic imaging must be modeled, although this has been previously done for VLT/SPHERE \cite{Ygouf2013A&A...551A.138Y, Paul2014A&A...572A..32P}. The removal of the reference star would significantly reduce observing time if post-processing techniques like ours can be informed by other wavefront sensing observations. Many future missions plan to incorporate WFS information into their high-contrast imaging observations, and WFE optimization routines will still be necessary to improve the modeling of the PSF due to changes from WFE drift. This technique is broadly applicable across many observing modes allowing for observer flexibility based on target characteristics, and with its proven efficacy will increase the efficiency of high-contrast imaging programs.

\section{Conclusion} \label{sec:conclusion}
In this paper, we implemented our post-processing technique that utilizes JWST WFS information as priors to estimate the WFE of an observation with synthetic PSF models while simultaneously searching for a companion. This analysis is performed on JWST/NIRISS AMI and FP imaging simulations. For the first time, we have extracted the posterior distributions for both the WFE and companion parameters simultaneously using pyMultiNest. With this technique, we found: 

1) Our estimate of the WFE of the target observation produces a model that achieves higher contrast sensitivity compared to the model produced from the WFE taken directly from the simulated WFSC data representing the typical two day WFE estimate provided for JWST. In the AMI case, our WFE estimate improves the modeling of the PSF over all separations sampled, while in the full pupil case it improves the PSF model for separations $<$ 0.455''.

2) This technique achieves contrasts close to the photon noise limit in the AMI case. This is in part due to the reduced number of degeneracies between primary mirror segments from seven sub-apertures as opposed to the 18 mirror segments in the full pupil case. However, this achievement is also due to the extended structure of the AMI PSF, more sensitive to small changes in the Hexike coefficients.

3) The derived sensitivity limit of our technique is less than a factor of two times the derived photon noise limit. Although evaluated in a different plane than our technique, the traditional interferometric technique appears sensitive to companions at roughly twice the derived photon noise limit when there is $\sim$ 5nm WFE drift from the calibrator star, displaying the comparable performances of both techniques.

4) We have proven the ability to recover a faint, close companion with a contrast of 8 magnitudes in the F430M filter and a separation of 0.2055'' (1.5$\lambda$/D, 3.16 NIRISS pixels). The recovery of the companion parameters performs comparably to the traditional interferometric post-processing techniques.

5) Our approach shows the ability to perform high contrast imaging without a reference star observation, significantly reducing the total allocated time for such a program.

\acknowledgments % equivalent to \section*{ACKNOWLEDGMENTS}       

This research made use of POPPY, an open-source optical propagation Python package originally developed for the James Webb Space Telescope project \cite{Perrin2012SPIE.8442E..3DP}.

M.D.F benefited from JPL's Strategic University Research Partnership (SURP). This work was supported by NASA through the JWST NIRCam project through contract number NAS5-02195 (M. Rieke, University of Arizona, PI). This research was carried out at the Jet Propulsion Laboratory, California Institute of Technology, under a contract with the National Aeronautics and Space Administration. M.D.F acknowledges S.F.

% References
\bibliography{report} % bibliography data in report.bib
\bibliographystyle{spiebib} % makes bibtex use spiebib.bst

\end{document}